# Real-Time-Data Analytics in Raw Materials Handling

## Christopher Rothschedl, Roland Ritt, Paul O'Leary, Matthew Harker, Michael Habacher, Michael Brandner

Chair of Automation, University of Leoben, Leoben, Austria.
(e-mail: roland.ritt@unileoben.ac.at).

# 1  Introduction

This paper proposes a system for the ingestion and analysis of real-time sensor and actor data of bulk materials handling plants and machinery. It references issues that concern mining sensor data in cyber physical systems (CPS) as addressed in O'Leary et al. [2015].

The advance of cyber physical systems has created a significant change in the architecture of sensor and actor data. It affects the complexity of the observed systems in general, the number of signals being processed, the spatial distribution of the signal sources on a machine or plant and the global availability of the data. There are different definitions for what constitutes cyber physical systems Baheti and Gill [2011], Geisberger and Broy [2012], IOSB [2013], Lee [2008], NIST [2012], Park et al. [2012], Spath et al. [2013a,b], Tabuada [2006]: the most succinct and pertinent to the work shown in this paper is the definition given by the IEEE Baheti and Gill [2011] and ACM[1]:

> A CPS is a system with a coupling of the cyber aspects of computing and communications with the physical aspects of dynamics and engineering that *must abide by the laws of physics*. This includes sensor networks, real-time and hybrid systems.

Results computed from sensor and actor data **must** obey the equations used for modelling the physics of the observed system — this fundamentally poses an *inverse problem*. Such problems are not covered sufficiently by literature addressing *mining of sensor data*, see for example Esling and Agon [2012], Fuchs et al. [2010], Keogh and Kasetty [2003], Last et al. [2004]. Even available standard books, such as Aggarwal [2013] on mining sensor data, do not discuss the special nature of sensor data. Typically, present approaches of mining data rely on correlation as being a sole, reliable measure for significance. It is not taken into account that the inverse solutions to the model-describing equations are required to establish a semantic link between a sensor observation and its precedent cause. Without this link — without *causality* — there can be no physics based knowledge discovery.

---

[1] ACM/IEEE International Conference on Cyber-Physical Systems (ICCPS) (iccps.acm.org)

The underlying data analytics problem can be described generally by the following statements:

1. The momentum of what is called Industry 4.0 promotes an increasing amount and availability of data. A suitable data ingestion system becomes necessary to acquire real-time sensor and actor data on a global scale. The fundamental concept on how to acquire, transport, ingest, and provide data needs to be sufficiently secure and adaptable enough to accommodate data of mining machines that may be located in remote areas.
2. Mathematical tasks are required to apply data analytics to industrial data sets, such as the solution of inverse problems and optimal-control-type problems.
3. Complex systems are modelled mathematically by following principles gained from modelling simple engineering systems, e.g., a vibrating string or a vibrating beam. These can be modelled using differential equations, ordinary and partial. More sophisticated mathematical models will be required to conquer the expanding complexity of modern mechatronic systems.
4. Data analytics will determine the particular causes to specific behaviour witnessed by sensor and actor data. Inverse problems are fundamental to accomplish such tasks. Additional metadata is required to accurately interpret the results of inverse models, as inverse problems do not have unique solutions per definition.
5. Extracting knowledge from data lies beyond simple information extraction. A more profound view on the philosophy of science points towards the necessity of assigning semantic information to data channels to establish such investigations. The metaphorical parallels between machine behaviour and natural language provide a form of knowledge extraction. It can be shown that machines have their own specific polysyllabic language. Once identified, it can be efficiently queried for symbolic patterns of normal or anomalous behaviour.

## 2    System Premiss

As an extension of Ackoff's work (Ackoff [1989]), Embrechts (Embrechts et al. [2005]) proposes the data mining pyramid consisting of the terms *data*, *information*, *knowledge*, *understanding* and *wisdom*. Embrechts does not provide any definitions for these terms, Ackoff offers intuitive but rather nebulous definitions; both do not provide a scientific basis for mining sensor data. Based on the integral idea we propose the fundamental concept behind the data analytics in Fig. .

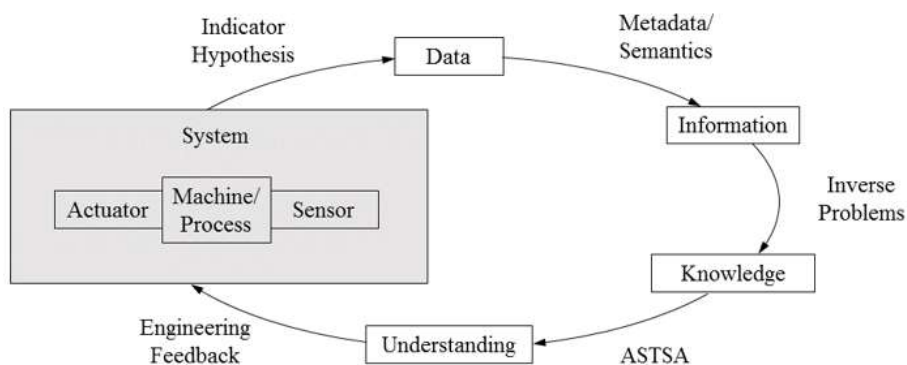

Fig. 1: The process behind the data analysis system.

The presented hierarchy illustrates how the questions of processing large data sets can be approached in a coherent and structured manner. The fundamental relationships of this premiss are:

1. A suitable indicator hypothesis builds the basis for the collection of data. If a specific sensor is chosen, an implicit indicator hypothesis has been selected as well, i.e., a temperature sensor defines that temperature is of relevance for the task.
2. Once acquired, data is only present as a simple stream of numbers; *metadata* adds meaning to the data. Beyond that, *context* is required to establish *significance*: a temperature value can have entirely contrasting significances for measurements of two different sources.
3. System models and the solution of the corresponding inverse problems are required to establish a causal link between measurement data and its possible cause. In general, there are no unique solutions to inverse problems.
4. Hence, a-priori knowledge is necessary to find the desired solution. These results of the inverse problems (the causes) constitute *knowledge*.
5. Effects of *human-machine interaction* must be considered to gain *understanding* of the whole system behaviour. Our approach, *Advanced Symbolic Time Series Analysis (ASTSA)*, is based on the emergence of language as it is modelled by the philosophy of phenomenology. The basic principle consists of symbols that are assigned to actions — *verbs*. The symbols for states are *nouns*. *Adverbs* and *adjectives* are used to predicate the verbs and nouns. *Punctuation* represents different lengths of pauses. Following such a segmentation, the time series is automatically converted into a sequence of symbols, enabling symbolic querying.
6. The whole process serves the understanding of what was originally only a stream of numbers. *Engineering feedback* can be derived from understanding the system response behaviour to certain loads and circumstances. Existing systems can be optimised and future revisions benefit from this as well.

## 3     Data Ingestion

A versatile data handling system is necessary to conquer large sets of time series data in a structured and efficient manner. Before such a system is able to provide any data, it has to ingest data following a specific workflow. In the course of the ingestion process, data is collected, quality-checked, and merged with corresponding metadata before it is prepared to fit a consistent data model. Sensor values are handled in the same way as derived measurements, i.e., the force of a hydraulic cylinder calculated from its dimensions (metadata) and its pressure values (time series from sensors).

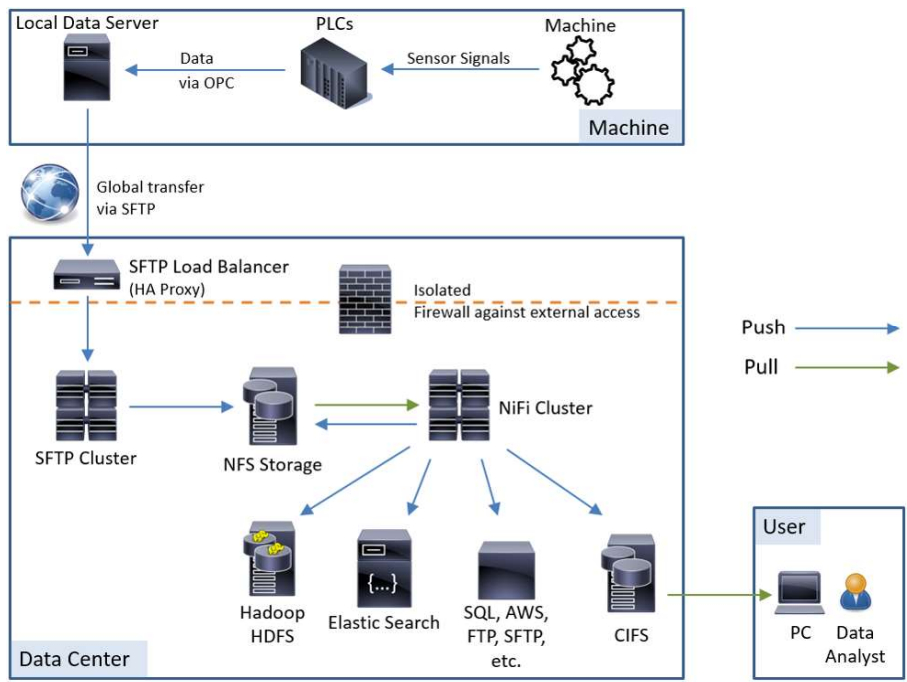

Fig. 2:   This illustration shows the main processes of data ingestion. The top section corresponds to the machine or plant on which data is being collected, while the bottom part represents the data center located at a different location. The data is provided in several formats after it has been ingested.

The concept describing the data ingestion process is illustrated in Fig. 2. Data of a machine's sensors is collected from its main programmable logic controller (PLC) and stored on a local data server before it is transported via a secured connection to the data center. After passing quality control, the data is stored permanently according to the data model and specified data manipulation workflows can be triggered on the cluster. Ultimately, the data is made available to consumers (data analysts, report recipients, domain experts, etc.) in different formats: this ensures that all users are independent in their choice of working environment.

The data is stored as a contiguous data stream as a result of the data ingestion process, see Fig. 3. The data input can be split, e.g., as daily exports of a buffering database running on the local data server at the machine's location. The data of all packets are merged to a contiguous, multi-channel stream of time series. When a user requests data from the system, they have the experience of querying the machine directly and in real-time. This opens the door to evaluations spanning time ranges varying from days to months and years. Furthermore, time ranges fitting a machine's operation characteristics can be queried, such as the time for loading a vessel in the case of analysing a ship loader. This permits a complete differentiation between input and output segmentation.

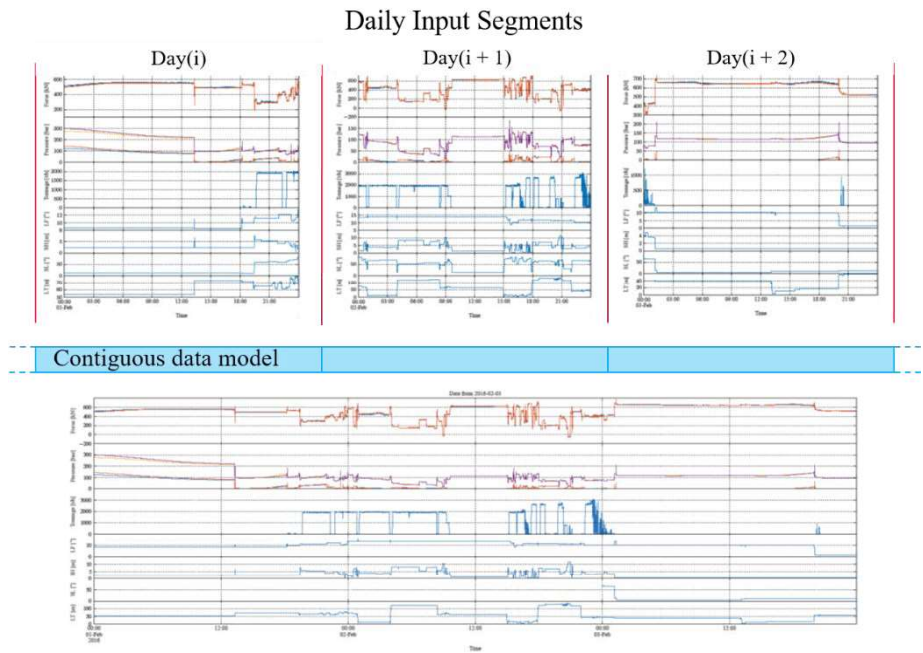

Fig. 3:  Three single days of data are assembled to a contiguous data stream. The illustrated contiguous section corresponds to the time portion a ship loader needs to load a vessel: this enables evaluations based on time ranges that are significant to particular fields of interest.

## 4    Systems Currently Being Monitored

Four mining machines that are currently being monitored using the approach presented in this paper are shown in Fig. 4. Data of these systems is collected constantly with a sampling interval of 1s. Typically, 50 to 850 sensor signals are collected, depending on the complexity of the monitored system.

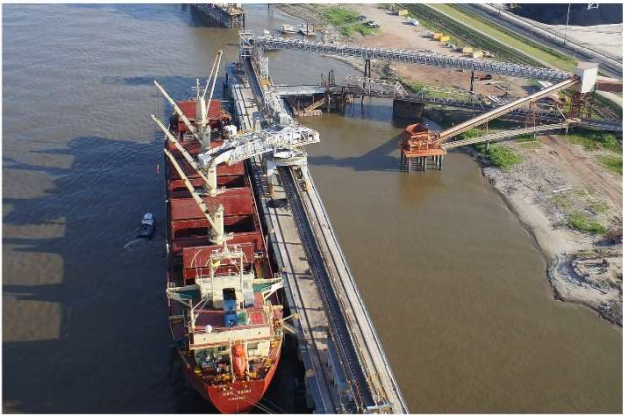
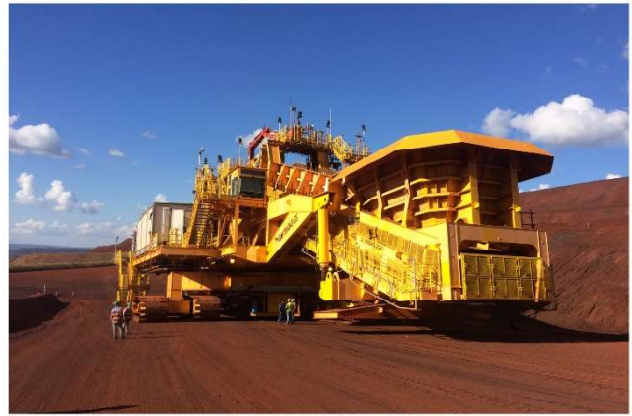
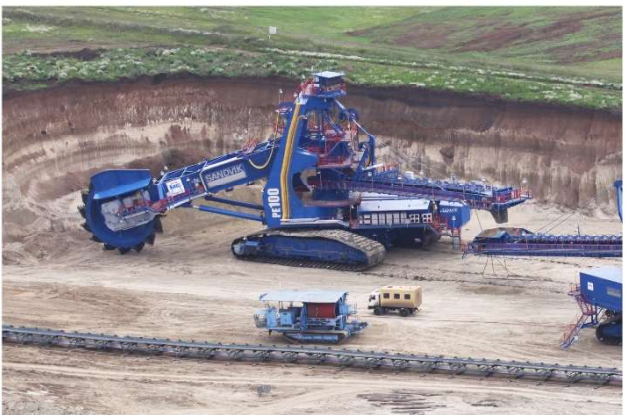
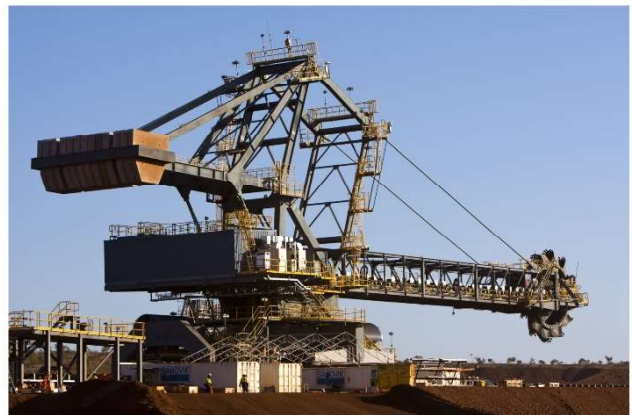

Fig. 4:    Examples of four systems that are currently being monitored using the described approach: a) ship loader, b) mobile sizing rig, c) bucket-wheel excavator, d) bucket-wheel reclaimer. The sensor channels of these systems are monitored with a sampling interval of 1s. (Sources: (a) – http://www.flickriver.com/photos/impalatermnals_images/17557941415/, retrieved on 2016-02-08; (b), (c), (d) – Courtesy of Sandvik.)

## 5    Exemplary Data Evaluations

The collection and analysis of data can be used for many different aspects of evaluating a machine during its life-cycle:

**Condition Monitoring:** Undoubtedly, data analytics can be used to address questions regarding condition monitoring or preventative maintenance, see Rothschedl [2016]. However, in this work we focus on issues that have received less attention in literature, e.g., incident analysis.

**Commissioning:** If data is already collected during the commissioning phase of a machine, analysing it can support shortening the time needed for this phase. Controlled tests can be verified with manageable effort and unexpected response behaviour to specific load scenarios can be detected. On several occasions, it was possible to identify sensors that delivered erroneous values for only a few samples a day. Judging from the nature of such error patterns, it would not be possible for a commissioning engineer to spot these defective sensors without such a system.

**Fleet management:** Insights gained from analysing one machine can support understanding the behaviour of other machines of similar design. For example, two identical bucket-wheel excavators were monitored which are operated in the same mine, handling the same type of material. The characteristics of both machines matched in many aspects. In contrast, two similar ship loaders exhibited behaviour that was significantly different. This raises the question whether these machines fulfil the conditions required to be ergodic systems.

**Automatic Operations Recognition:** With ASTSA, several data channels can be combined to define machine states. Sequences of these states refer to corresponding operation modes which can be used to characterise how a machine is being controlled. These sequences support the identification of inappropriate operations that may lead to damages or to missing performance goals.

**Incident Analysis:** Incidents with equipment in mining environments bear serious financial and legal issues. Unplanned maintenance and repair work in such environments and locations quickly reach immense financial dimensions, also because associated materials handling processes are interrupted, provoking serious follow-up costs. Liability for injury and damages are the main concerns from the legal point of view. The analysis of real-time operational data prior to incidents supports the determination of the possible causes for their occurrences and, hence, can provide more certainty to the financial and legal claims. Although this form of analysis can shed light on the clarification of far-reaching issues, this topic has been rarely mentioned in literature. It is evident that incident analysis plays a major role when working with mining machines.

**Logistics Optimisation:** The analysis of long-term time series allows evaluations based on aggregated data: the distribution of conveyed material over the full slewing range of a machine over a long period of time can support identifying unevenly distributed component utilisation. Such problems can often be avoided or mitigated if the logistics of a machine are adapted.

Two exemplary evaluations are presented:

## 5.1    Incident Analysis

The figures below (Fig. 5 and Fig. 6) show the results of performing incident analysis for a bucket-wheel excavator. The analysis shows a large number of events distributed over time and conspicuous times during which no events occurred: this is with most certainty operator-dependent behaviour of the system as a whole.

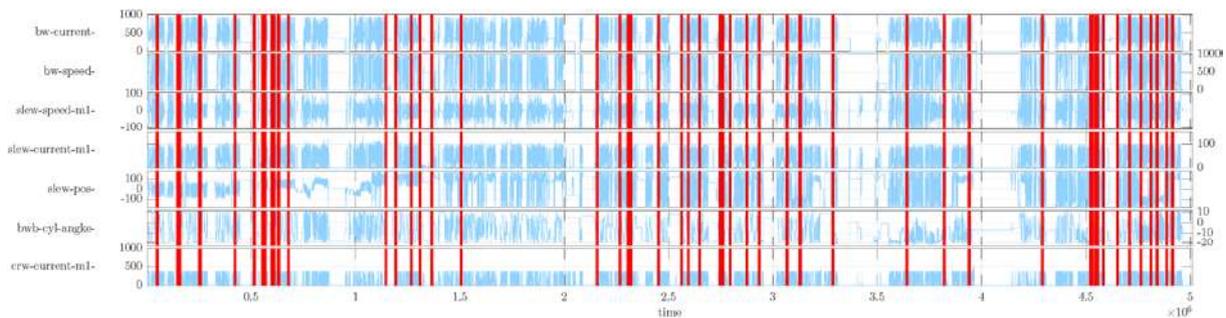

Fig. 5:    This example of incident analysis shows data for a time period of two months, acquired with a sampling time of 1s. Each vertical line corresponds to an event; 63 events were found in total by using Advanced Symbolic Time Series Analysis (ASTSA). Every event corresponds to an inappropriate operation of the machine: the data can be zoomed in on automatically for every single event to perform local analysis, i.e., in the seconds and minutes right before the occurrence of the event (see Fig. 6).

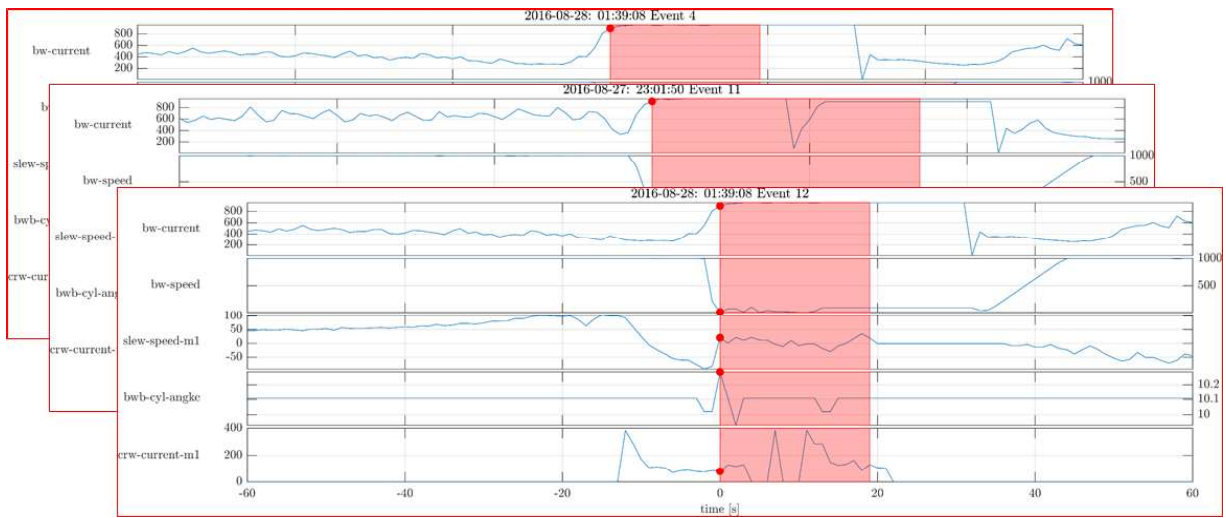

Fig. 6:    Plots of the identified events with 1s resolution for three of the 63 events reported in Fig. 5.

## 5.2    Long-Term Logistics Optimisation

The data shown in Fig. 7 is the polar histogram of loading on the slew bearing of a bucket-wheel reclaimer. The data has been aggregated with $t_s$=1s over an observation period of one year. Interestingly, the overloading in one quadrant is not visible on a daily basis. The higher loading, evident from aggregated long-term data in the figure, has significant consequences on the life span of the bearing.

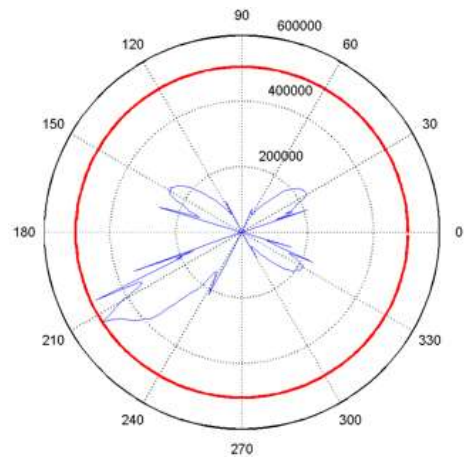

Fig. 7:   Polar histogram of loading on the slew bearing of a bucket-wheel reclaimer. The data has been aggregated with a sampling time of 1s over an observation period of one year.

## 6   Conclusions

The collection of very large real-time data series from plant and machinery is highly relevant in a mining context. A strongly structured approach is required, if the best use is to be made of the data. The results are relevant for both, machine constructors and also their operators. It is significantly more than just preventative maintenance.